\documentclass[11pt,a4paper,preprintnumbers,notitlepage,nofootinbib]{revtex4-1}

\usepackage[T1]{fontenc} 
\usepackage{graphicx}
\usepackage{xspace}
\usepackage{amsmath,amssymb,mathtools}
\RequirePackage{flushend}
\RequirePackage[colorlinks,citecolor=blue,urlcolor=blue,linkcolor=blue]{hyperref}
\usepackage[utf8]{inputenc}
\usepackage{color}
\usepackage{natbib}
\usepackage{enumitem}
\usepackage{mathtools}
\DeclarePairedDelimiter{\abs}{\lvert}{\rvert}

\renewcommand{\Re}{{\rm Re}}
\renewcommand{\Im}{{\rm Im}}
\newcommand{\ti}[1]{\tilde{#1}}

\newcommand{\beq}{\begin{eqnarray}}
\newcommand{\eeq}{\end{eqnarray}}
\newcommand{\be}{\begin{equation}}
\newcommand{\ee}{\end{equation}}

\begin{document}

\preprint{}

\title{Precision Predictions for Charged Higgs Boson Decays in the
  Real and Complex NMSSM \\\textit{{\mdseries{\small Talk presented at the International Workshop on Future Linear Colliders (LCWS2021),\hspace{3cm} 15-18 March 2021, Online Conference, C21-03-15.1.}}}}

\author{
Thi Nhung Dao$^{1\,}$\footnote{E-mail: \texttt{dtnhung@ifirse.icise.vn}},
Margarete M\"{u}hlleitner$^{2\,}$\footnote{E-mail:
	\texttt{margarete.muehlleitner@kit.edu}},
Shruti Patel$^{2,3\,}$\footnote{E-mail: \texttt{shruti.patel@kit.edu}},
Kodai Sakurai$^{2\,}$\footnote{Speaker}\;\footnote{Address after April 2021, Department of Physics, Tohoku University, Sendai, Miyagi 980-8578, Japan; E-mail: \texttt{kodai.sakurai.e3@tohoku.ac.jp}}
\\[9mm]
{\small\it
$^1$Institute For Interdisciplinary Research in Science and Education, ICISE,}\\
{\small\it 590000 Quy Nhon, Vietnam.}\\[3mm]
{\small\it
$^2$Institute for Theoretical Physics, Karlsruhe Institute of Technology,} \\
{\small\it Wolfgang-Gaede-Str. 1, 76131 Karlsruhe, Germany.}\\[3mm]
{\small\it$^3$Institute for Nuclear Physics, Karlsruhe Institute of Technology,
76344 Karlsruhe, Germany.}\\[3mm]
}
\begin{abstract}

%In light of the current situation that a new particle has not been discovered, indirect searches through observables for Higgs bosons are rather important. This requires accurate theoretical predictions for such observables in order to compare them with the precision measurements in experiments. 
We present the full next-to-leading order (NLO) supersymmetric (SUSY)
electroweak and SUSY-QCD corrections to the decay widths of the
charged Higgs boson decays into on-shell final states in the framework
of the CP-conserving and CP-violating Next-to-Minimal Supersymmetric
Model (NMSSM) of Ref.~\cite{Dao:2020dfb}. 
The newly calculated corrections have been implemented in the code
{\tt NMSSMCALCEW}.  
In these proceedings, we discuss the impact of the NLO corrections on
the charged Higgs boson branching ratios in a wide range of the parameter
space that is still compatible with the experimental constraints.  
We also investigate the effect of CP violation in these corrections. 
\end{abstract}

\maketitle

\newpage
\section{Introduction}

The Standard Model (SM) has been established as a low-energy effective theory that describes electroweak symmetry breaking (EWSB). 
However, there exist several unsolved phenomena,  such as neutrino oscillations, dark matter, baryon asymmetry of the Universe, and inflation.  
Since they are not explained within the framework of the SM, new physics is needed to address these puzzles. 
%Also,  motivated by theoretical problems of the SM, like e.g.~the
%hierarchy problem and the strong CP problem, various new paradigms and new mechanisms have been proposed. 
One among the models beyond the SM that solve the hierarchy problem is
given by supersymmetry (SUSY)
~\cite{Golfand:1971iw, Volkov:1973ix, Wess:1974tw, Fayet:1974pd, Fayet:1977yc, Fayet:1976cr, Nilles:1983ge, Haber:1984rc, Sohnius:1985qm,Gunion:1984yn, Gunion:1986nh}.  
Its simplest version is realized by the Minimal Supersymmetric SM
(MSSM), where two Higgs doublet fields are required in order to
ensure supersymmetry and  to keep the theory anomaly-free. 
In this work, we focus on the Next-to minimal Supersymmetric SM
(NMSSM)~\cite{Fayet:1974pd,Barbieri:1982eh,Dine:1981rt,Nilles:1982dy,Frere:1983ag,Derendinger:1983bz,Ellis:1988er,Drees:1988fc,Ellwanger:1993xa,Ellwanger:1995ru,Ellwanger:1996gw,Elliott:1994ht,King:1995vk,Franke:1995tc,Maniatis:2009re,Ellwanger:2009dp}. Besides
the virtue of solving the so-called $\mu $ problem~\cite{Kim:1983dt},
the NMSSM disposes of an interesting Higgs boson phenomenology.  
Its Higgs sector contains a complex singlet field in
addition to the two Higgs doublet fields which yields three CP-even
Higgs bosons, two CP-odd Higgs bosons and two charged Higgs bosons
after EWSB, implying interesting Higgs physics full of variety.
Furthermore, the Higgs sector can be CP-violating already at the tree
level unlike the MSSM.  

So far the LHC experiments have not found any direct evidence of new
particles beyond the SM, leading to lower bounds on the masses of new particles. 
In addition, measurements of the observables for the discovered Higgs
boson (e.g.~production cross sections, decay branching ratios and
couplings) are consistent with the predictions of the SM within the
current experimental uncertainties, meaning that the discovered Higgs
boson behaves very SM-like.  
%By these observations, parameter space of the NMSSM is seviararly constrained.  
While the SUSY particles are required to be heavy enough to comply
with the constraints from the direct searches they may still give
significant contributions in the higher-order corrections to Higgs
observables. In addition, the properties of the discovered Higgs boson
will be measured very precisely in future collider experiments, such
as the high-luminosity (HL-) LHC~\cite{ApollinariG.:2017ojx}, the
International Linear Collider (ILC)~\cite{Baer:2013cma}, the Future
Circular Collider (FCC-ee)~\cite{Gomez-Ceballos:2013zzn}, or the
Circular Electron Positron Collider
(CEPC)~\cite{CEPC-SPPCStudyGroup:2015csa}. Hence, precise theoretical
calculations are inevitable in order to be able to identify indirect signs of new physics. 
In the context of the NMSSM, higher-order corrections have been
calculated to the NMSSM Higgs boson masses up to two-loop accuracy
both in the CP-conserving and CP-violating NMSSM. For a recent
overview, see \cite{Slavich:2020zjv}.
For the Higgs boson decays, full one-loop corrections to neutral Higgs
boson decays into two gauge bosons and two fermions have been
calculated in Ref.~\cite{Domingo:2018uim} in the CP-violating NMSSM. The effects of Sudakov logarithms on fermionic decays of heavy Higgs
bosons have been studied in Ref.~\cite{Domingo:2019vit}.
For Higgs-to-Higgs decays, not only the one-loop
corrections~\cite{Nhung:2013lpa, Belanger:2017rgu} but also the two
loop corrections with $\mathcal{O}(\alpha_{S}
\alpha_{t})$~\cite{Muhlleitner:2015dua} have been studied. 
Furthermore, in Ref.~\cite{Baglio:2019nlc}, all on-shell two-body
decays of neutral Higgs bosons have been evaluated at NLO electroweak
(EW) and NLO SUSY-QCD order in the NMSSM with CP violation.
In Ref.~\cite{Dao:2019nxi}, the NLO EW corrections to charged Higgs boson
decays into a $W$ boson and a neutral Higgs boson have been calculated
and the gauge dependence, which arises from higher-order corrections
to the mass and due to the resummed $Z$ factor of the external Higgs boson 
has been studied. This problem and possible strategies to preserve or
restore gauge invariance has been further investigated in \cite{Domingo:2020wiy}. 
In Ref.~\cite{Dao:2020dfb}, we completed the computation of the full
NLO corrections to the on-shell two-body decays of the charged Higgs
bosons in the NMSSM with CP violation. We evaluated the NLO SUSY-EW
and SUSY-QCD corrections to the decays of the charged Higgs bosons
into quarks, leptons, electroweakinos, squarks and sleptons for the
first time~\footnote{In the previous implementation in the program {\tt
    NMSSMCALC}~\cite{Baglio:2013iia}, the charged Higgs decay widths
  included the state-of-the art QCD
  corrections to the decays into quarks and squarks as well as the resummed SUSY-QCD corrections and SUSY-EW corrections for the decays into quarks and leptons.}. 
These new NLO corrections have been implemented in the new version of
the program {\tt NMSSMCALCEW}~\cite{Baglio:2019nlc}~\footnote{The
  program is available at the URL: \url{https://www.itp.kit.edu/~maggie/NMSSMCALCEW/}}. 
In these proceedings, we summarize our results obtained in
Ref.~\cite{Dao:2020dfb}, and discuss the sizes of the pure NLO
corrections for the charged Higgs boson decays into quarks and
electroweakinos. We furthermore show numerical results for the effects
of CP violation on the loop-corrections to the decays into quarks. 

\section{The Lagrangian of the NMSSM}

We briefly describe the Higgs potential of the NMSSM and the electroweakino sector in order to fix our conventions and input parameters. 
We work in the framework of the scale-invariant NMSSM with $Z_{3}$
symmetry. The tree-level Higgs potential obtained from the $F$- and
$D$-terms of the supersymmetric Lagrangian and the soft SUSY-breaking
Lagrangian is given by 
\begin{align}\label{eq:potential}
\notag
V_{H}&=|\lambda S|^2\left(H_{u}^\dagger H_u+H_{d}^\dagger H_d\right)
       +\left|-\epsilon_{ij}\lambda\left(H_d^iH_u^j\right)+\kappa S^2\right|^2
        \\ \notag
&+\frac{1}{2}g_2^2\left|H_u^\dagger
  H_d\right|^2+\frac{1}{8}(g_1^2+g_2^2)\left(H_u^\dagger
  H_u-H_d^\dagger H_d\right)^2 \\  
&+m_{H_u}^2H_u^\dagger H_u+m_{H_d}^2H_d^\dagger
  H_d+m_S^2\left|S\right|^2+
  \left[
 -\epsilon_{ij}\lambda A_\lambda\left(H_d^iH_u^j\right)S
  +\frac{1}{3}\kappa A_\kappa S^3+{\rm h.c.}\right], 
\end{align}
where $\lambda$ and $\kappa$ are complex dimensionless parameters
defined in the superpotential, and the corresponding terms
proportional to $A_{\lambda}$ and $A_{\kappa}$ arise from the soft SUSY-breaking terms. 
 The gauge couplings of the $U(1)_{Y}$ and $SU(2)_{L}$ symmetry are
 denoted by $g_{1}$ and $g_{2}$, respectively. 
The Higgs doublet fields $H_{u}$ and $H_{d}$ and the complex singlet
field $S$ are expressed in terms of the component fields and vacuum
expectation values (VEVs) as
\begin{align}
H_u=
e^{i\varphi_u}\begin{pmatrix}
h_u^+\\
\frac{1}{\sqrt{2}}(v_u+h_u+ia_u)
\end{pmatrix},\ 
H_d=
\begin{pmatrix}
\frac{1}{\sqrt{2}}(v_d+h_d+ia_d) \\
h_d^-
\end{pmatrix},\ 
S=\frac{1}{\sqrt{2}}e^{i\varphi_s}(v_s+h_s+ia_s) \;, \label{eq:HiggsVEVexpand}
\end{align}
where $v_u, v_d $ and $v_s$ are the VEVs of $H_u$, $H_d$, and $S$, respectively. 
{The EW VEV is given by $v^{2}=v_{u}^{2}+v_{d}^{2}\simeq (246~{\rm GeV})^{2}$. }
The two CP-violating phases $\varphi_u$ and $\varphi_s$ denote the phase
differences between the VEVs. The neutral component fields are
transformed into the mass eigenstates through the orthogonal rotation matrix $\mathcal R$,
\begin{align}
(h_1,h_2,h_3,h_4,h_5,G^0)^T=  \mathcal R (h_d, h_u,
h_s, a_d, a_u, a_s)^T, \label{eq:rotgaugemasstree}
\end{align}
where $G^0$ is the neutral Nambu-Goldstone (NG) boson. Similarly, the mass
eigenstates for the charged Higgs bosons are obtained by,
\begin{align}
\begin{pmatrix}
G^\pm\\H^\pm
\end{pmatrix}
=
\begin{pmatrix}
-c_\beta & s_\beta \\
s_\beta & c_\beta
\end{pmatrix}
\begin{pmatrix}
h_d^\pm\\h_u^\pm
\end{pmatrix},
\end{align}
where the mixing angle $\beta$ is defined by $\tan\beta=v_{u}/v_{d}$. 

The tree-level Higgs sector of the CP-violating NMSSM is described by
eighteen independent input parameters which we choose as
\begin{align}
& m_{H_d}^2, m_{H_u}^2, m_S^2,M_W^2, M_Z^2, e, 
  \tan\beta, v_s, \varphi_s, \varphi_u, 
 \abs{\lambda}, \varphi_\lambda, \abs{\kappa},
\varphi_\kappa, \Re A_{\lambda},{\Im A_\lambda}, \Re A_{\kappa},
{\Im A_\kappa}~. \label{eq:inputHiggs}
\end{align}
Here the three Lagrangian parameters $g_1$, $g_2$ and $v$ have been 
replaced by the three physical observables $M_W$, $M_Z$ and the
electric coupling $e$. The effective $\mu$ parameter can be expressed
as
\beq
\mu_{\rm eff} =
\abs{\mu_{\rm eff}}e^{i\varphi_\mu} =
  \frac{|\lambda| v_s }{\sqrt{2}} e^{i(\varphi_\lambda + \varphi_s)}
  \;.
\eeq 
{The real part of $A_\lambda$, $\Re A_{\lambda}$, can be replaced by the charged Higgs boson mass $M^{2}_{H^{\pm}}$ through 
\begin{align}
M^{2}_{H^{\pm}}=M_{W}^{2}+\frac{\lambda v_{S}}{\sin(2\beta)}\left(\sqrt{2} \Re A_{\lambda}+\kappa v_{S}\right)-\frac{\lambda^{2} v^{2}}{2}~.
\end{align}
The soft SUSY-breaking parameters $m_{H_d}^2, m_{H_u}^2, m_S^2,\Im A_{\lambda}, \Im A_{\kappa}$ are fixed by the tadpole conditions for the five neutral Higgs bosons. }

Adding the superchiral singlet field $\hat{S}$ introduces an
additional degree of freedom, the singlino $\tilde{S}$, in the electroweakino sector. 
This results in five neutralino mass eigenstates, which are related to
the gauge eigenstates via the 5$\times$5 unitary matrix $N$, 
\begin{align}
\chi^{0}=N\psi^{0} \;,
\end{align}
with
$\psi^{0}=(\tilde{B},~\tilde{W}^{3},~\tilde{H}_{d},~\tilde{H}_{u},~\tilde{S})^{T}$,
where $\tilde{B},~\tilde{W}^{3}$ stand for neutral gaugino states,
$\tilde{S}$ for the singlino state and
$~\tilde{H}_{d}$, and $\tilde{H}_{u}$ are the Higgsino states. 
The chargino mass eigenstates in the basis of the Weyl spinors for the charginos, $\ti{\chi}^+_L=(\ti{\chi}^+_{L_1},\ti{\chi}^+_{L_2})^T,\
 \ti{\chi}^-_R=(\ti{\chi}^-_{R_1},\ti{\chi}^-_{R_2})^T$, are obtained
 by rotating the spinors in gauge basis, $\psi^-_R=(\tilde{W}^-, \tilde{H}^-_d)^T,\
\psi^+_L=(\tilde{W}^+, \tilde{H}^+_u)^T$, as
\begin{align}
 \ti{\chi}_L^+=V\psi_L^+,\ \ \ \ti{\chi}_R^-=U\psi_R^- \;, 
 \end{align}
with the 2$\times$2 unitary matrices $U$, $V$. 
The specific expressions for the mass matrices of the neutralinos and
charginos in terms of the input parameters can be found in Eqs.~(17)
and (21) of Ref.~\cite{Dao:2020dfb}. By definition, they are
diagonalized by the unitary matrices $N$, $U$ and $V$.

\section{The Charged Higgs Boson Decay Widths Including Higher-Order
  Corrections} 
In Ref.~\cite{Dao:2020dfb}, we evaluated the NLO SUSY-EW corrections
and SUSY-QCD corrections to the two-body on-shell decay widths of the
charged Higgs bosons, i.e., $H^{+}\to t\bar{b}$, $H^{+}\to \nu \bar{\tau}$, $H^{+}\to \chi_{i}^{+} \chi_{j}^{0}$ ($i=1,2,~j=1,...5$), $H^{+}\to\ti{t}\ti{b}$ and $H^{+}\to \ti{\tau}\ti{\nu}$. 
In order to get UV finite results, we used a mixed OS and
$\overline{\rm DR}$ renormalization scheme for the Higgs sector and
the OS scheme for the gauge sector and the SM fermion sector.  
For the renormalization of the electroweakino, the squark and
the slepton sectors, both the OS scheme and the $\overline{\rm DR}$
scheme were utilized. 
In this section, we give schematic formulae for the partial decay
widths into $t\bar{b}$ and electroweakinos, including the higher-order corrections. 

For the decay $H^{+}\to t\bar{b}$, the loop-corrected partial decay
width can schematically be written as
\begin{align}\label{eq:htotbNLO}
\Gamma(H^{+}\to t\bar{b})^{\rm NLO}
=\Gamma^{\rm LO~imp.}_{H^{+}\to t\bar{b}}+\Gamma^{\rm SUSYQCD}_{H^{+}\to t\bar{b}}+\Gamma^{\rm SUSYEW}_{H^{+}\to t\bar{b}}. 
\end{align}
%+\Gamma^{\rm H^{+}G^{-}/W}_{H^{+}\to t\bar{b}}+\Gamma^{}_{H^{+}\to t\bar{b} \gamma}
The improved leading-order (LO) contribution $\Gamma^{\rm
  LO~imp.}_{H^{+}\to t\bar{b}}$, which already includes the
state-of-art QCD corrections and the resummed SUSY-QCD and SUSY-EW
corrections, is given {in terms of $\mu_{t}=m_{t}^{2}/M_{H^{\pm}}^{2}$ and $\mu_{b}=m_{b}^{2}/M_{H^{\pm}}^{2}$    } by 
\begin{eqnarray}
\Gamma^{\rm LO~imp.}_{H^+\rightarrow t\bar{b}} &=& \frac{3 G_F
M^3_{H^\pm}}{4\sqrt{2}\pi}
|V_{tb}|^2 \, \beta^{1/2}\left(\mu_{t}, \mu_{b}\right) \, \left\{ (1-\mu_{t} -\mu_{b}) \left[\mu_{t}\frac{1}{ \tan^2\beta } \left( 1+ \frac{4}{3} \frac{\alpha_s}{\pi} \delta_{tb}^+ \right)\right. \right. \nonumber\\
&& \left. \left. +\mu_{b} \tan^2 \beta R^2  \left( 1+ \frac{4}{3}\frac{\alpha_s}{\pi} \delta_{bt}^+ \right) \right]-4\mu_{t}\mu_{b} R \left( 1+ \frac{4}{3}\frac{\alpha_s}{\pi} \delta_{tb}^- \right) \right\} \label{eq:htotb} \;,
\end{eqnarray}
with
\be 
\beta^{1/2}(x,y) =(1-x-y)^2-4xy,
\ee
and where $G_F$ denotes the Fermi constant, $\alpha_s$ the strong
coupling constant, and $V_{tb}$ the top-bottom CKM matrix element. 
Explicit expressions for the QCD correction factors 
$\delta_{tb}^+,\delta_{bt}^+$ and $\delta_{tb}^-$ can be found in 
 Ref.~\cite{Spira:2016ztx}. 
The universal SUSY-QCD and SUSY-EW corrections that are enhanced in
the large $\tan\beta$ regime are resummed into effective bottom Yukawa
couplings. These $\Delta_b$ corrections
~\cite{Hall:1993gn,Hempfling:1993kv,Carena:1994bv,Pierce:1996zz,Carena:1998gk,Carena:1999py,Carena:2002bb,Guasch:2003cv,Spira:2016ztx}
are included in the decay width through the $R$ factor, $R =
\frac{1}{1+\Delta_b} \left[ 1 - \frac{\Delta_b}{\tan^2\beta}
\right]$. The explicit formula for $\Delta_b$ is
given in Ref.~\cite{Baglio:2013iia}.  
The contributions $\Gamma^{\rm SUSYQCD}_{H^{+}\to t\bar{b}}$ and
$\Gamma^{\rm SUSYEW}_{H^{+}\to t\bar{b}}$ correspond to one-loop
SUSY-QCD and SUSY-EW corrections, respectively, where subtraction
terms are added to avoid double counting arising from the 
$\Delta_{b}$ corrections.  
 The SUSY-EW corrections contain, besides the pure vertex correction,
 the external leg corrections from the mixing between the charged
 Higgs boson and the charged NG boson and the $W$ boson, respectively,
 and the real photon emission $\Gamma_{H^+ \to t\bar{b}\gamma}$.  
Specific expressions for these NLO corrections are given in Ref.~\cite{Dao:2020dfb}. 
  
The partial decay widths for the decays into electroweakinos are written in the same notation as
\begin{align}\label{eq:NLOwidthEWino}
\Gamma(H^{+}\to \chi_{i}^{+} \chi_{j}^{0})^{\rm NLO}
=\Gamma^{\rm LO }_{H^{+}\to \chi_{i}^{+} \chi_{j}^{0}}+\Gamma^{\rm SUSYEW}_{H^{+}\to \chi_{i}^{+} \chi_{j}^{0}},  
\end{align}
$(i=1,2,~ j=1,...5)$. 
%+\Gamma^{\rm H^{+}G^{-}/W}_{H^{+}\to \chi_{i}^{+} \chi_{j}^{0}}+\Gamma^{}_{H^{+}\to \chi_{i}^{+} \chi_{j}^{0} \gamma}
The LO decay width is given by
\beq
\Gamma^{\rm LO}_{H^+ \to \ti{\chi}_i^0\ti{\chi}^+_j } &= &
\frac{\beta^{1/2}\left(\mu_i,\mu_j\right)M_{H^\pm}}{16\pi}\Bigg[ \left(1-\mu_i-\mu_j \right) 
\left(\abs{g_{ H^+ \tilde{\chi}_i^0 \tilde{\chi}^{-}_j }^{L, {\rm tree}}}^2+\abs{{g}_{ H^+ \tilde{\chi}_i^0 \tilde{\chi}^{-}_j }^{R, {\rm tree}}}^2\right) \\
&&-2 \sqrt{\mu_i\mu_j}\left({g}_{ H^+ \tilde{\chi}_i^0 \tilde{\chi}^{-}_j }^{L,{\rm tree}}(g_{ H^+ \tilde{\chi}_i^0 \tilde{\chi}^{-}_j }^{R,{\rm tree}})^* +{g}_{ H^+ \tilde{\chi}_i^0 \tilde{\chi}^{-}_j }^{R,{\rm tree}}(g_{ H^+ \tilde{\chi}_i^0 \tilde{\chi}^{-}_j }^{L,{\rm tree}})^*\right)\Bigg]\,,\notag
\eeq
with $\mu_i = M_{\tilde{\chi}_i^0}^2/M_{H^\pm}^2$ and $\mu_j =
M_{\tilde{\chi}_j^+}^2/M_{H^\pm}^2$, 
where the tree-level couplings are given by
\begin{align}
g_{H^+  \ti{\chi}_i^0\ti{\chi}^-_j}^{R,{\rm tree}}&=
-\frac{1}{\sqrt{2}}{V}_{j2}\Big((g_1N_{i1}+g_2N_{i2})c_{\beta}e^{i \varphi_u}+\sqrt{2}\lambda^\ast N_{i5}s_{\beta}\Big) 
-g_2 e^{i \varphi_u} {V}_{j1}N_{i4} c_{\beta} \,,  
\\ 
g_{H^+  \ti{\chi}_i^0\ti{\chi}^-_j}^{L,{\rm tree}}&=
-\frac{1}{\sqrt{2}}{U}_{j2}^\ast \Big({-}g_1N_{i1}^\ast s_{\beta} {-}g_2N_{i2}^\ast s_{\beta} +\sqrt{2} e^{i \varphi_u}\lambda N_{i5}^\ast c_{\beta}\Big) 
-g_2{U}_{j1}^\ast N_{i3}^\ast s_{\beta}. \label{eq:lhcoupling}
\end{align}
There are no NLO SUSY-QCD corrections, and the NLO SUSY-EW
corrections comprise besides the vertex corrections the mixing between
$H^{+}$ and the charged Goldstone and $W$ bosons, respectively, and
the real photon emission $\Gamma^{}_{H^{+}\to
  \chi_{i}^{+}\chi_{j}^{0}\gamma }$. The concrete formulae are also
given in Ref.~\cite{Dao:2020dfb}.  

\section{Numerical results}
\subsection{Branching ratios for $H^{+}$ at LO}
%--------------------------
\begin{figure}[t]
 \centering
 \includegraphics[width=.55\textwidth]{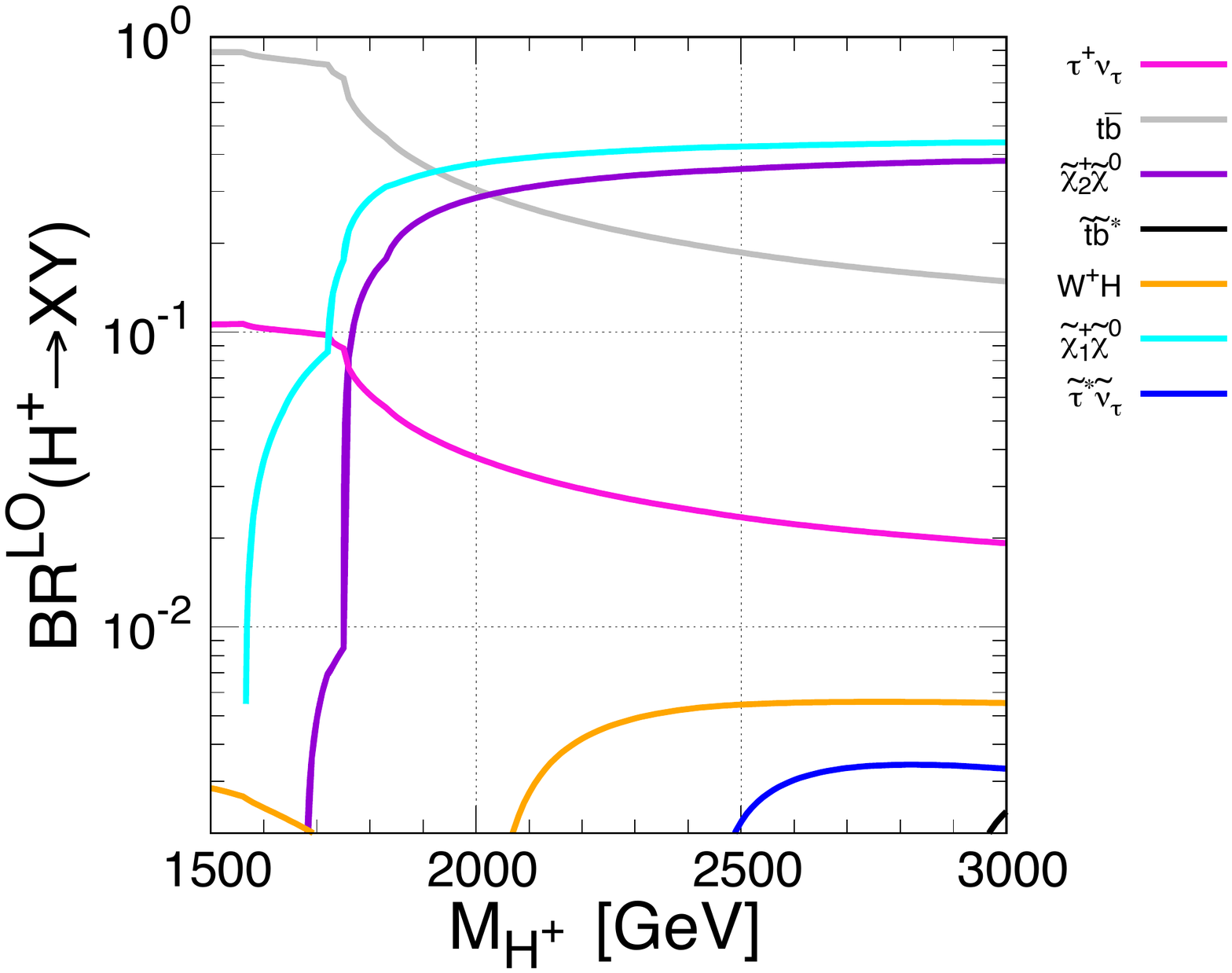}
 \vspace{-1.5cm}
 \caption{The branching ratios of the charged Higgs boson $H^+$ decays
   $H^{+}\to \tau^{+} \nu_{\tau}$ (pink), $H^{+}\to t \bar{b}$ (gray),
   $H^{+}\to WH$ (orange), $H^{+}\to \ti{\chi}^{+}_{1}\ti{\chi}^{0}$
   (cyan), $H^{+}\to \ti{\chi}^{+}_{2}\ti{\chi}^{0}$ ({violet}),
   $H^{+}\to \ti{t}_{1}\ti{b}_{1}^{\ast}$ (black), $H^{+}\to
   \ti{\tau}\ti{\nu_{\tau}}^{\ast}$ (blue), at LO as a function of the charged Higgs boson
   mass $M_{H^{\pm}}$.  
. }
 \label{FIG:LOBR}
\end{figure}
%--------------------------
We show the branching ratios of the charged Higgs bosons at LO in
Fig.~\ref{FIG:LOBR} to illustrate the behavior as a function of the
charged Higgs boson mass. 
In the plot, pure NLO corrections are switched off for all decay processes. 
The input parameters for the Higgs  sector and the electroweakino sector are fixed as 
\beq
\label{eq:BPCP}
\begin{array}{llllllllllll}
 \tan\beta&=&10.14 \;, & \;
                                                       |\lambda|&=&0.093 \;,
& \;  |\kappa|&=&-0.0821 \;, \\ 
M_{H^{+}}&=&1500\  {\rm GeV},\;& \mu_{\rm eff}&=&-891\ {\rm GeV},& \;  \Re A_\kappa&=&- 1.6\ {\rm TeV}, \\
|M_1|&=&752 \ {\rm GeV}, & \; |M_2|&=& 806 \ {\rm GeV}.  
\end{array} \nonumber
\eeq
The soft SUSY-breaking parameters for the sfermions and the gluino,
and the complex phases are fixed as
 \begin{align}
%\notag
&|M_3|=2334 \
{\rm GeV}, \; |A_t|= 3.7\ {\rm TeV},  \; |A_b|= 2\ {\rm TeV}, \;  |A_\tau|=
                                                                      2\ {\rm TeV}, \notag \\   
& m_{\tilde{Q_{3}}}=1.43\ {\rm TeV}, \; m_{\tilde{t}_R}=1.93\ {\rm TeV},  \;  m_{\tilde{b}_R}=2.16\ {\rm
  TeV}, \; m_{\tilde{L}_3}=1.22\ {\rm TeV}, \; \notag\\ 
& 
m_{\tilde{\tau}_R}=1.21\ {\rm TeV}, \; m_{\tilde{u}_R,\tilde{c}_R}=m_{\tilde{d}_R,\tilde{s}_R}=m_{\tilde{Q}_{1,2}}=m_{\tilde{L}_{1,2}}=m_{\tilde{e}_R,\tilde{\mu}_R}=3\ {\rm TeV}, \nonumber \\
&\varphi_{M_{1},M_{2},M_{3}}=\varphi_{A_{t},A_{b},A_{\tau}}=\varphi_{\mu}=\varphi_{\kappa}=0 \;.
\end{align}
 In this plot, the decays into the neutral Higgs bosons plus the $W$
 boson and the electroweakinos, respectively, are summed  with respect
 to the final states as 
 \begin{align}
\mbox{BR}(H^+ \to W^+ H) &\equiv \sum_{i=1}^{3} \mbox{BR}(H^+ \to W^+
H_i) + \sum_{j=1}^{2} \mbox{BR}(H^+ \to W^+ A_j) \;, \notag \\
\mbox{BR}(H^+ \to \tilde{\chi}^+_{1,2} \tilde{\chi}^0) &\equiv \sum_{i=1}^5
\mbox{BR}(H^+ \to \tilde{\chi}^+_{1,2} \tilde{\chi}^0_i) \;.
\end{align} 
Similarly,  the decays into the slepton final states are summed up. 
The mass spectrum for this bench mark point is given by
 \begin{align}
&M_{H_1} =123.33 \mbox{ GeV}, 
\; M_{H_2}= 1.09 \mbox{ TeV}, \; M_{H_3}= M_{A_1}=1.50 \mbox{ TeV}, \; \nonumber \\
&M_{A_2}= 1.93 \mbox{ TeV} \;, 
M_{\chi^{+}_{1}}=817.87~\mbox{ GeV}, \;
M_{\chi^{+}_{2}}=927.87~\mbox{ GeV}, \; \notag\\
&M_{\chi^{0}_{1}}=747.69~{\rm GeV}, \; 
M_{\chi^{0}_{2}}=820.07~{\rm GeV}, \; 
M_{\chi^{0}_{3}}=894.20~{\rm GeV}, \; \notag\\
&M_{\chi^{0}_{4}}=929.24~{\rm GeV}, \; 
M_{\chi^{0}_{5}}=1.56~{\rm TeV}, \; 
M_{\ti{t}_{1}}=1.38~{\rm TeV}, \; \notag\\
& \; 
M_{\ti{b}_{1}}=1.43~{\rm TeV},  
M_{\ti{\tau}_{1}}=  
M_{\ti{\tau}_{2}}= 
M_{\ti{\nu}_{\tau} }=1.2~{\rm TeV} \;.
\end{align} 
As can be inferred from the plot, for $M_{H^{+}}\lesssim1.9\  {\rm
  TeV}$, the decay into the top-bottom final state is the main decay mode because of the $\tan\beta$ enhancement for the bottom Yukawa coupling. 
Most of the decay channels into the electroweakinos are kinematically
open for $1.9 \lesssim M_{H^{+}}\  {\rm TeV}$, so that the
branching ratios into electoroweakinos become more important than
those into top-bottom.

\subsection{Impact of the NLO Corrections on the Charged Higgs Branching Ratios}
%--------------------------
\begin{figure}[t]
 \centering
 \includegraphics[width=.48\textwidth]{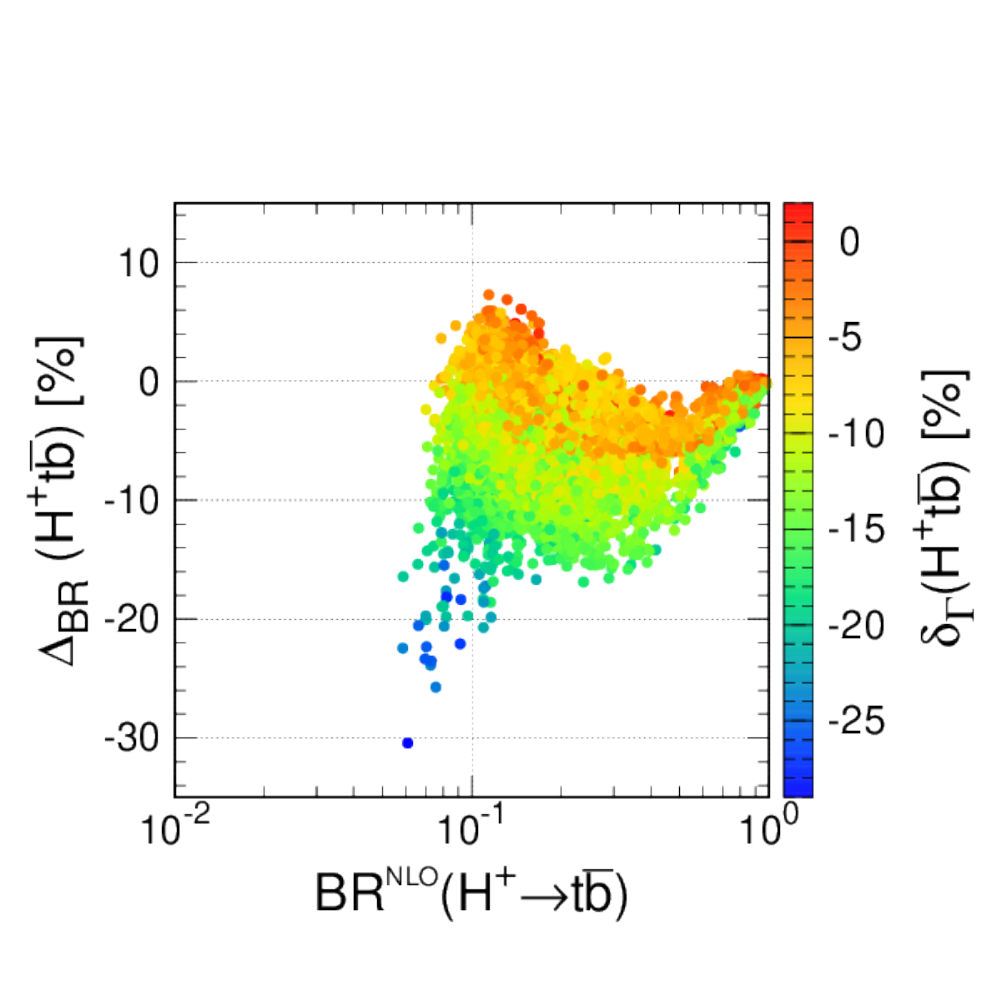}
  \includegraphics[width=.48\textwidth]{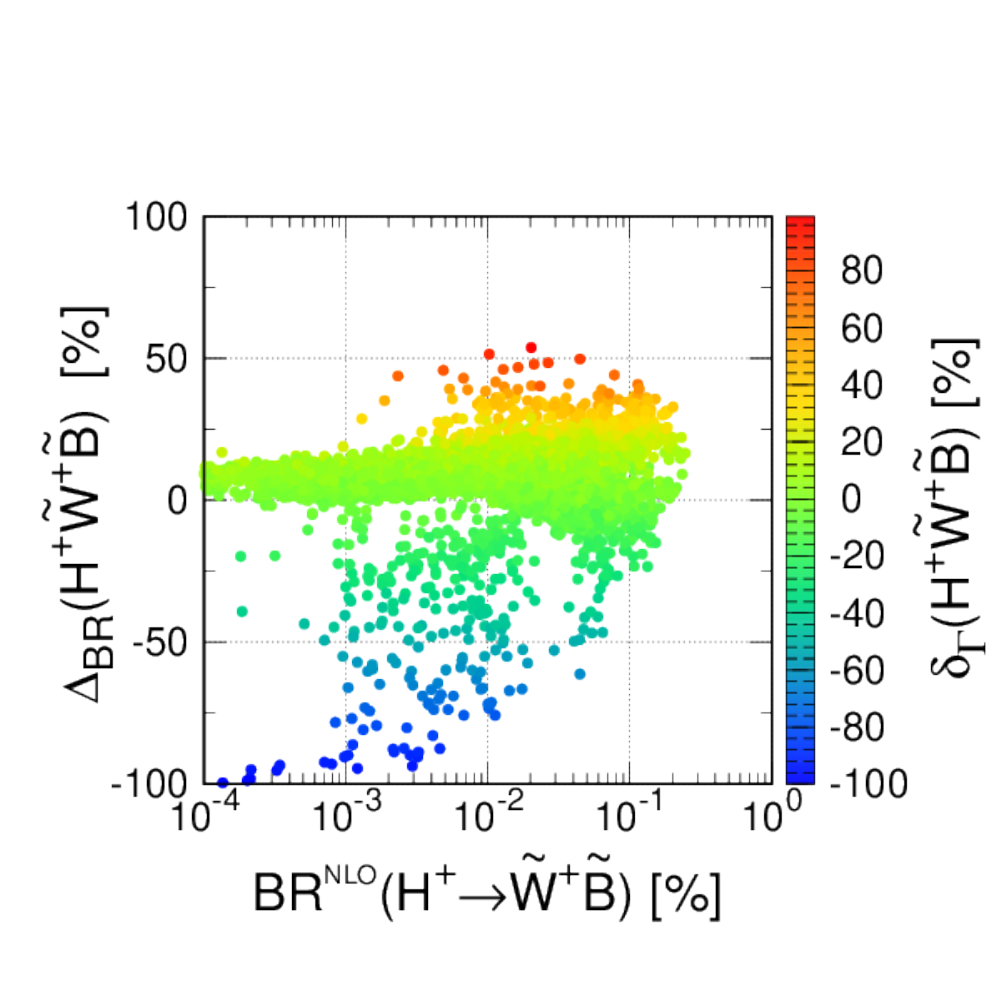}
% \vspace{1.0cm}
\caption{Correlation between the relative change for the decay
  branching ratio, $\Delta_{\rm BR}$, and the NLO branching ratio for
  $H^{+}\to t\bar{b}$ (left panel) and $H^{+}\to \ti{W}^{+}\ti{B}$
  (right panel).  
The color bar shows the size of the relative NLO corrections for the decay
width $\delta_{\Gamma}$.  
}
 \label{FIG:NLOBR}
\end{figure}
%--------------------------
In this section, we discuss the impact of the NLO corrections on the
branching ratios of the charged Higgs boson for the decays into the
top and bottom quark pair and the electroweakino final states. 
In order to describe the size of the NLO corrections for the branching
ratios as well as the partial widths of the decays $H^{+}\to
XY$, we introduce the following quantities, 

\begin{align}\label{eq:DeltaBR}
\Delta_{\rm BR}^{\rm }(H^{+}X Y)&=\frac{{\rm BR}^{\rm NLO} (H^{+}\to X Y) 
-{\rm BR}^{\rm LO}(H^{+}\to X Y)}
{ {\rm max}( {\rm BR}^{\rm NLO} (H^{+} \to X Y),{\rm BR}^{\rm
  LO}(H^{+} \to X Y) ) } \;,  \\
  \label{eq:Deltawid}
\delta_{\Gamma}(H^{+}X Y)&=\frac{\Gamma(H^{+}\to X Y)^{\rm
                                NLO} }{\Gamma(H^{+}\to X Y)^{{\rm
                                LO}} }-1   \;.
\end{align}
The relative change $\Delta_{\rm BR}$ is normalized to the maximum of
the branching ratio at NLO and LO, respectively, to avoid an
enhancement due to the smallness of ${\rm BR}^{\rm LO}$. 
The notation 'LO' basically means that we refer to the old implementation
in  {\tt NMSSMCALC} without the genuine SUSY-EW and
SUSY-QCD vertex corrections. This means that the LO quantities are
calculated with 'Higgs effective tree-level couplings', meaning that
the Higgs tree-level rotation matrix elements have been replaced by
the loop-corrected ones. Furthermore, these 'LO'
quantities also include the QCD corrections
and resummed SUSY-EW and SUSY-QCD corrections in effective quark
couplings as already implemented in the first release of {\tt
  NMSSMCALC} \cite{Baglio:2013iia}. 
Hence, the relative changes $\Delta_{\rm BR}$ and $\delta_{\Gamma}$
quantify the effect of the pure NLO corrections that were newly
evaluated in Ref.~\cite{Dao:2020dfb}.  
For the following scatter plots, we scanned the NMSSM input parameters
in the ranges given in Table.~1 together with the parameter settings
in Eqs.~(128) and (129) of Ref.~\cite{Dao:2020dfb} to understand the
typical size of the NLO corrections in a wide range of the parameter
space. We only retained points that are compatible with the LHC Higgs
data. For details, see Ref.~\cite{Dao:2020dfb}.
For the calculation of the NLO corrections, we used the OS1
renormalization scheme for the eletroweakino sector, the $\overline{\rm DR}$ scheme for the stop and sbottom sector, and the OS scheme for the stau sector. 
The definition of each scheme can be found in Ref.~\cite{Baglio:2019nlc}

In the left panel of Fig.~\ref{FIG:NLOBR}, we show the relative change
$\Delta_{\rm BR}$ for $H^{+}\to t\bar{b}$ as a function of the
branching ratio at NLO. The color code denotes the relative change of
the decay width. 
As written in Eq.~\eqref{eq:htotbNLO}, the total NLO corrections for
the partial width are determined by the sum of the SUSY-QCD and the
SUSY-EW corrections.   
The latter gives negative corrections ranging between $-15\%$ and
$-2\%$, and the former remains between $-15\%$ and $7\%$ so that the
relative change $\delta_{\Gamma}$ is negative  in most of the
parameter region and the maximum size $\abs{\delta_{\Gamma}}$ is
$29\%$. For this decay channel, the relative change of the branching
ratio $\Delta_{\rm BR}$ is basically correlated with that of the partial decay width. 

In the right panel of Fig.~\ref{FIG:NLOBR}, the relative change
$\Delta_{\rm BR}$ for the decay into the charged Wino and the Bino
final state, $H^{+}\to \ti{W}^{+}\ti{B}$, is shown as a function of
the corresponding branching ratio at NLO.  
We select the charged Wino-like chargino and Bino-like neutralino
states from the electroweakino mass eigenstates. 
They are identified by the square of the mixing matrix elements,
i.e. $|U_{i1}|^{2}$ and $|N_{j1}|^{2}$, which are required to exceed
0.5 for $\ti{W}^{+}$ and $\ti{B}$, respectively.  
In the plot, we applied cuts for the left tree-level coupling of
Eq.~\eqref{eq:lhcoupling} as well as for the mass difference among the
electroweakinos, because we found that artificially enhanced NLO
corrections can appear  without such a cut, which are discussed in
Sec. 5.5 of Ref.~\cite{Dao:2020dfb}.  
As can be inferred from the plot, the relative change of the branching
ratio is $\abs{\Delta_{\rm BR}}<50\%$ for the bulk of the parameter points. 
On the other hand, if the size of the branching ratio at NLO becomes
smaller than about $5\times 10^{-2}$, outliers appear and the relative
change $\Delta_{\rm BR}$ can reach almost $-100\%$.  
These large corrections arise from large contributions due to the wave
function renormalization constants (WFRCs) for electroweakinos, which
are involved in $\Gamma^{\rm SUSYEW}_{H^{+}\to \chi_{i}^{+}
  \chi_{j}^{0}}$ in Eq.~\eqref{eq:NLOwidthEWino}. 
The WFRCs contains mixing contributions between another electroweakino
state and the one in question, and they can be significant if the
other electroweakino has a large coupling with the charged
Higgs boson.  

%--------------------------
\begin{figure}[t]
 \centering
 \includegraphics[width=.42\textwidth]{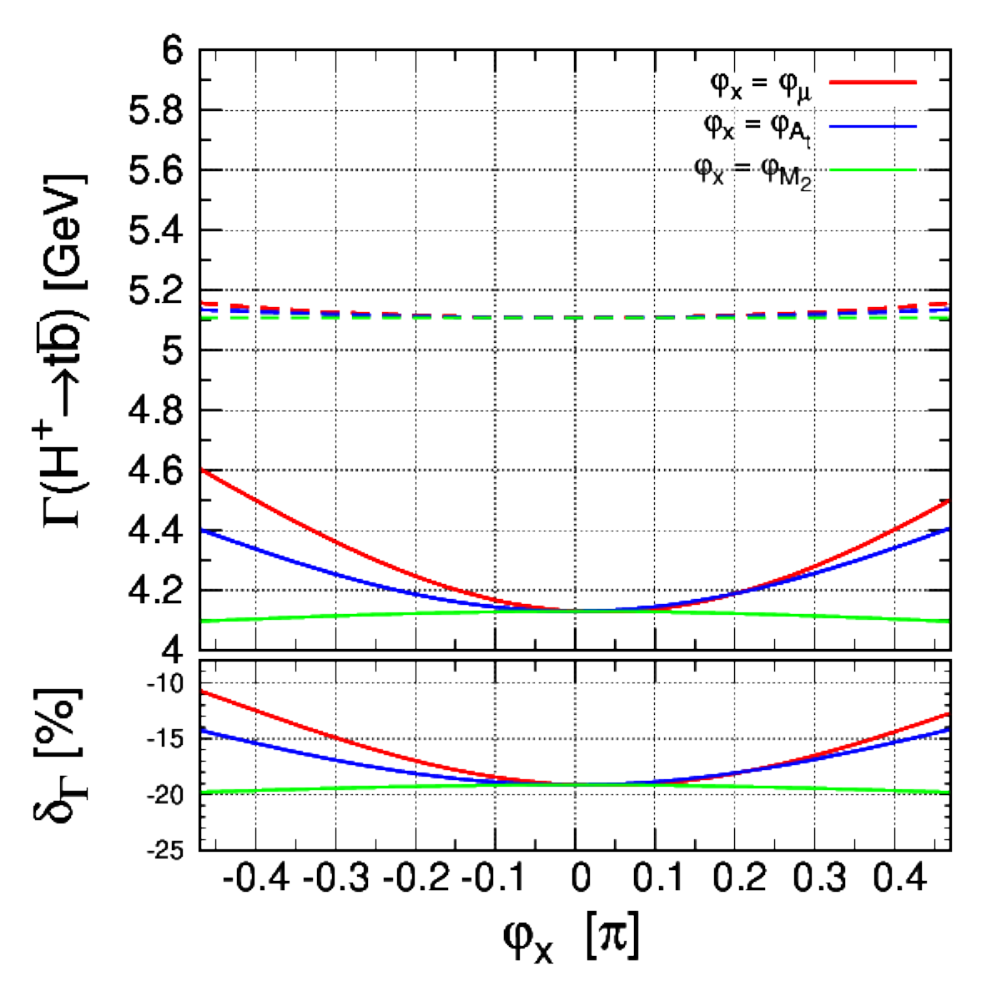}
\caption{
CP-violating phase dependence of the partial decay width for $H^{+}\to
t\bar{b}$ at LO (dashed lines) and NLO (solid lines) for the
CP-violating phases of $\mu_{\rm eff}$ (red), 
$A_{t}$ (blue) and $M_{2}$ (green).
The lower panel displays the dependence of the relative change of the
partial decay width $\delta _{\Gamma}$.  
 }
 \label{FIG:CPVhptotb}
\end{figure}
%--------------------------
Before we close this section, we discuss how CP-violating phases can
affect the NLO corrections to the decays of the charged Higgs
boson. We just examine the theoretical behavior of the relative change
of the partial width when the phase parameters are varied, and do not
take into account experimental constraints by the measurement of the
electric dipole moment (EDM) in this analysis.  
In order to illustrate the effect of CP violation, we show the partial
decay width for $H^{+}\to t\bar{b}$ as a function of complex phases
for $\mu_{\rm eff}$, $A_{t}$ and $M_{2}$ in Fig.~\ref{FIG:CPVhptotb}.  
In the plot, the benchmark scenario for the CP-conserving case
presented in Sec.~5.7 of Ref.~\cite{Dao:2020dfb} is used, and the
three phases are varied individually. 
Note that we set the phase $\varphi_\lambda$ such that we do not
encounter CP violation at tree level.
The nevertheless observed subtle phase dependence on $\varphi_\mu$ and
$\varphi_{A_{t}}$, respectively, originates from SUSY loop
contributions that are resummed in the $\Delta_{b}$ corrections. 
Once the pure NLO corrections are included, we see a stronger
dependence on the CP-violating phases. 
The relative change in the partial decay width can reach e.g. around $-10\%$ at
$\varphi_{\mu}\simeq -0.46$ while it is around -20\% at $\varphi_\mu=0$.  

\section{Conclusion}
We have summarised the results for the NLO SUSY-QCD and NLO SUSY-EW
corrections for various on-shell two body decays of the charged Higgs
boson, presented in Ref.~\cite{Dao:2020dfb}.  
The newly evaluated corrections are included in the latest version of
{\tt NMSSMCALCEW}.  
The NLO corrections are evaluated by using a mixed OS and $\overline{\rm DR}$ renormalization scheme in the Higgs sector. 
For the results presented here, in the electroweakino sector the OS1 scheme is used while the $\overline{\rm DR}$ (OS) scheme  is chosen for the squarks (slepton) sector. 
We analysed the relative change of the branching ratios  and the
partial decay widths, which are defined in Eqs.~\eqref{eq:DeltaBR} and
\eqref{eq:Deltawid}, for $H^{+}\to t\bar{b}$ and $H^{+}\to
\ti{W}^{+}\ti{B}$ in a wide range of the parameter space.  
We found that the NLO corrections to $ {\rm BR}(H^{+}\to t\bar{b})$ are moderate and can reach almost $-30\%$ at ${\rm BR}^{\rm NLO} \simeq 6\times 10^{-2}$. 
On the other hand, we found that for $H^{+}\to \ti{W}^{+}\ti{B}$,
negative large corrections can appear in a corner of the parameter space where the size of the branching ratio at NLO becomes small. 
We also presented the effects of CP-violating phases on the NLO corrections to
$H^{+}\to t\bar{b}$.  
We found that  the changes of the CP-violating phases can
significantly affect the NLO corrections to the decays of the charged
Higgs boson.  

\vspace{-.5cm}
\subsection*{Acknowledgments}\vspace{-.5cm}
In this work, T.N.D has been funded by the Vietnam National Foundation for Science
and Technology Development (NAFOSTED) under grant number 103.01-2020.17.  
M.M. and K.S. acknowledge support by the Deutsche Forschungsgemeinschaft
(DFG, German Research Foundation) under grant  396021762 - TRR
257. 

%%%%%%%%%%%%%%%%%%%%%%%%%%%%%%%%%%%%%%%

%%%%%%%%%%%%%%%%%%%%%%%%%%%%%%%%%%%%%%%

\end{document}